# A Mixed Initiative Framework for Semantic Web Service Discovery and Composition

**Jinghai Rao[a], Dimitar Dimitrov[b], Paul Hofmann[b] and Norman Sadeh[a]**

[a]School of Computer Science, Carnegie Mellon University, 5000 Forbes Avenue, Pittsburgh, PA, 15213 – USA.
{sadeh, jinghai}@cs.cmu.edu
[b]SAP, Dietmar-Hopp-Allee 16, D-69190 Walldorf, Germany.
{paul.hofmann, dimitar.dimitrov}@sap.com

**Abstract**

A central element of emerging Service Oriented Architectures (SOA) is the ability to develop new applications by composing enterprise functionality encapsulated in the form of services – whether within a given organization or across multiple ones. Semantic service annotations, including annotations of both functional and non-functional attributes, offer the prospect of facilitating this process and of producing higher quality solutions. A significant body of work in this area has aimed to fully automate this process, while assuming that all services already have rich and accurate annotations, In this article, we argue that this assumption is often unrealistic. Instead, we describe a mixed initiative framework for semantic web service discovery and composition that aims at flexibly interleaving human decision making and automated functionality in environments where annotations may be incomplete and even inconsistent. An initial version of this framework has been implemented in SAP's Guided Procedures, a key element of SAP's Enterperise Service Architecture (ESA).

## 1. Introduction

Service Oriented Architectures (SOAs) provide a framework within which enterprises expose functionality in the form of loosely coupled services that can be integrated and consolidated in response to demand for new applications (or services). Over the past several years, languages and frameworks have been proposed to develop and leverage rich semantic service annotations in support of both service discovery and composition functionality - e.g. [5,7,8]. A significant portion of this work has been devoted to scenarios aimed at automating service discovery and composition functionality (see surveys in [12,13]) – notable exceptions include efforts reported in [14,15,16]. While valuable, this work does not address the challenges involved in training personnel to efficiently and accurately develop the necessary service annotations and ontologies. Nor does it fully recognize the amount of effort involved in annotating legacy applications in use in both small and large organizations. What is needed for enterprises to be able to exploit the power of semantic web service technologies are tools that can effectively support their personnel from day one with significantly incomplete and possibly inconsistent annotations. These tools therefore need to be highly interactive in nature. They need to support users through suggestion, completion and verification functionality, while always allowing them to override their recommendations. In this article, we present a mixed initiative framework for semantic web service discovery and composition intended for that purpose. Mixed initiative functionality does not attempt to fully automate all decisions (e.g. [3,4]). Instead it is based on the premise that users should retain close control over many decisions while having the ability to selectively delegate tedious aspects of their tasks. Automated service discovery and composition functionality is merely used to selectively intervene and assist users in some of their tasks by providing suggestions, identifying inconsistencies, and completing some of the user's decisions. This enables the user to dynamically choose how much of the discovery and composition process to delegate and how much of it to retain control over.

The framework we present has been validated in the context of SAP's Guided Procedures, a key element of SAP's Enterprise Service Architecture (ESA) [17] and its Composite Application Framework (CAF) [18]. Specifically, CAF is built into SAP's NetWeaver [19] to support the development of cross-functional applications and business processes. Guided Procedures (GP) [20] is the tool developed as part of CAF to enable lay users (i.e. users without software development skills) to set up and execute new collaborative business processes out of existing functionality and applications. Target users include SAP personnel as well as SAP consultants and "analyst" users responsible for the customization, refinement and composition of applications and services at client organizations. It should be pointed out that the mixed initiative framework presented in this paper is not specific to GP and that it could be applied across a broad range of other service discovery and composition scenarios.

The remainder of this paper is organized as follows. Section 2 introduces Guided Procedures and our mixed initiative framework for semantic service discovery and composition. Section 3 details our modeling framework, Section 4 discusses the underlying semantic web reasoning and service discovery and composition functionality used in our framework. This includes the way in which some of this functionality has been broken down in support of our mixed initiative framework. Section 5 revisits the Guided Procedures scenario introduced earlier. Section 6 discusses the framework's current implementation and presents an initial set of empirical results. Concluding remarks are provided in Section 7.

## 2. A Mixed Initiative Framework for Service Discovery and Composition

SAP's Guided Procedures (GP) allow users to define new composite applications and processes by re-using, integrating and orchestrating existing functionality encapsulated in the form of composable elements. In GP, composable elements comprise primitive functionality ("callable objects" and "actions" in the GP jargon) as well as composite functionality (or "blocks" in the GP jargon), including complex processes and services.

Usage scenarios range from assisting SAP consultants as they tailor and combine existing SAP modules and functionality to capture the processes and policies of a particular company, to scenarios where analyst users build new composite applications that leverage functionality exposed by third party partners in the form of web services.

The mixed initiative framework for semantic web service discovery and composition described in this paper has been implemented as a recent addition to SAP's Guided Procedures, though its applicability extends beyond this particular environment. It enables users to annotate composable elements with semantic profiles that refer to concepts in an open collection of ontologies. These annotations are used by mixed initiative functionality to support users at design time as they specify abstract requests for composite applications, search for, select among, and compose available services to satisfy these requests. This functionality is presented to users in the form of simple services that can selectively intervene at different points in this often highly iterative process (Figure 1). They provide suggestions, offer to complete tedious steps and verify decisions made by users, while always allowing them to manually override their recommendations and regain control.

The development of composite services tends to be an iterative process, where users successively specify and relax requirements while tailoring and combining existing functionality to satisfy these requirements. The GP framework is intended to accommodate different work styles ranging from "top down" approaches, where a user specifies an abstract description of a desired composite service to more serendipitous or "bottom-up" approaches where users directly edit and compose existing services – and anything in between. Abstract descriptions of composite services will be in the form of constraints on desired input and output parameters as well as on the state of affairs both prior to and after invoking the composite service. A simple example of such an abstract description could read "I want a service that takes a RFQ as input and generates a Quote as output". A more complex description could be of the form "I want a service that takes care of all RFQs that have not yet been processed".

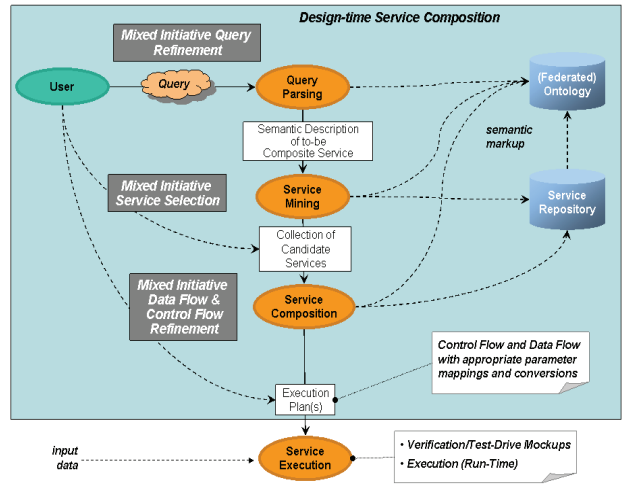

**Figure 1:** Simplified Workflow – Mixed initiative design-time functionality supports users as they refine the specification of composite services, identify and select relevant services and compose them. Actual workflows often involve multiple iterations.

While relying on semantic annotations to guide mixed initiative service discovery and composition, our framework recognizes that GP users cannot be expected to be experts in annotating composable functionality. Instead, it is understood that typical users will often fail to initially identify (or specify) relevant annotations. Accordingly our framework is designed to operate with partial annotations and help users become better at annotating composable functionality over time. As annotations become richer and more accurate, the quality of the guidance provided by our framework also improves and users gradually learn to take advantage of it. Because mixed initiative functionality is provided in an unobtrusive way, it never hinders users.

Broadly speaking, our framework's mixed initiative functionality consists of simple, often customizable, services capable of providing suggestions and feedback to users as they deal with each of the following three sets of key decisions:

1. Semantic Discovery: This functionality enables users to search repositories of composable functionality, based on both functional and non-functional attributes (e.g. [9]) – e.g. searching for one or more services that could help build a composite application. Functional attributes include input, output parameters as well as preconditions and effects. Non-functional attributes refer to other relevant characteristics such as accuracy, quality of service, price, owner, access control restrictions, etc.

2. Semantic Dataflow Consolidation: This functionality assists users by automatically suggesting ways of mapping input and output parameters of composable functionality elements as they are being composed. This includes functionality to automatically complete an existing step – this is similar to "code completion" functionality except that it is based on semantic reasoning. It also includes verification functionality that flags seemingly erroneous or inconsistent assignments.
3. Semantic Control Flow Consolidation: This is similar, except that here were are concerned with the order in which services will be executed. This includes reasoning about the availability of necessary input variables and, more generally, about the preconditions and effects associated with the execution of different services. Again this functionality can be provided in the form of suggestions or to help verify the correctness of decisions made by users. It can be invoked for an entire process or just for two or more steps in a process currently under construction. Suggestions can include the introduction or removal of particular sequencing constraints. It may also involve identifying and adding one or more additional steps to a partial process. In general, users should be able to specify how much they want to delegate at any point in time, e.g. whether to request help with a small subproblem or with a more extensive portion of an existing solution.

As users interact with the above functionality, they should always have the flexibility to selectively revise and complete existing annotations. Over time, we also envision adding global analysis functionality. This would go beyond just verifying the correctness of composite applications to include identifying redundancies and inefficiencies in processes.

## 3. Underlying Representation Model

Below, we briefly review the way in which ontologies are organized and provide an overview of the underlying service model and annotations used in our framework.

### Ontologies

An ontology is simply a description of concepts relevant to a given domain along with attributes/properties characterizing these concepts. By relying on shared ontologies, namely by agreeing on the definition of common concepts, developers within a given organization can define composable functionality elements that refer to the concepts in these ontologies. So can enterprises as they selectively expose composable functionality elements to business partners in the form of (semantic) web services.
Our Model builds on an extensible and customizable set of ontologies defined in W3C's OWL language [1]:

- **Core Ontologies**: These are ontologies that are used to capture the semantics of core composable functionality elements. They include:
  o **Core Composable Element Ontology** – this ontology is currently based on a simplified fragment of the OWL-S ontology [2] and includes the definition of concepts such as composable elements (aka services in OWL-S), input parameters, output parameters, preconditions, effects as well as basic properties to describe composite processes assembled from composable elements. Preconditions specify conditions that need to hold prior to invoking a composable element, whereas effects describe the possible outcomes associated with the element's invocation. A small fragment of the Core Composable Element ontology is depicted in Figure 2. The ontology can also be extended to include non-functional attributes - these can also be defined as part of domain-specific ontologies.
  o **Core GP Ontology** – this ontology builds on the Core Composable Element ontology to define GP specific concepts such as Callable Objects, Actions, Blocks, Input Parameters, Output Parameters, Parameter Types, Strings, Integers, Business Objects, etc. Together these definitions provide an explicit semantics of core GP concepts. In the future, similar Core Ontologies could be developed for other process development or service discovery and composition environments..
- **Domain-specific Ontologies** – this extensible library of domain-specific ontologies is expected to grow over time to include:
  o Generic SAP ontologies of concepts such as those found in generic business objects
  o Application and Sector specific ontologies, namely ontologies of concepts and policies particular to a given application or a given sector
  o Company-specific ontologies capturing company-specific concepts and policies.

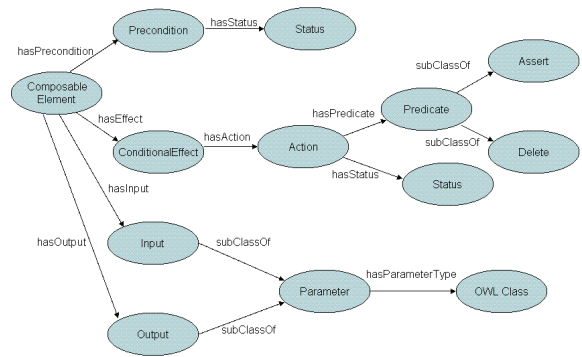

**Figure 2.** Composable Elements: profiles with inputs, outputs, preconditions and effects

## The Core Composable Element Ontology

A composable (functionality) element can be either an atomic service (e.g. a GP Callable Object, Action, including external services wrapped as such) or a composite service (e.g. a GP Block or Process). It is described (or "annotated") by its:

- Input parameters
- Output parameters
- Preconditions
- Effects
- Non-functional attributes

Both preconditions and effects are currently represented using "status" objects. The preconditions are currently interpreted as being part of a conjunction, namely all preconditions need to hold before activating the composable element. A composable element can have multiple conditional effects, each representing different mutually exclusive possible outcomes. In other words, the particular conditional effects that will hold following the execution of a composable element will depend on the actual execution of that component (e.g. whether a request is approved or rejected or whether execution of a service is successful or not). A conditional effect is itself a collection of actions, each either asserting or deleting status objects. Status objects are defined in relation to OWL classes. A status class can have several properties. For example, in describing a purchase order processing, service a "submitted" class can be used to indicate that a purchase order has been submitted. These properties are instantiated at runtime based on bindings (defined at design time) to relevant input and output parameters.

A composite process is described in terms of a process model, as illustrated in Figure 3. The model details both its control structure and data flow structure. A process is recursively defined as either a "Composable Element" or a "Composite Process". A "Composite Process" contains one or more sub-processes. Sub-processes are to be executed according to control constructs. Examples of control constructs include "sequence", "choice" and "parallel". Each process has a set of parameters, including "inputs", "outputs", "preconditions" and "effects". A "Composite Process" is also described in terms of "Perform" constructs that specify how data flows across the process. This is done using a "Consolidation" construct that maps input and output parameters of composable elements onto one another ("dataflow consolidation").

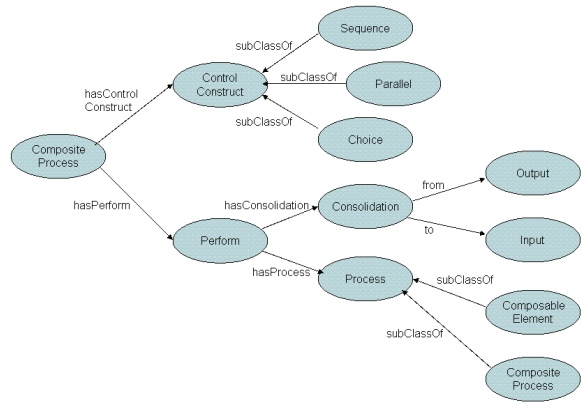

**Figure 3.** Simple process ontology.

## Annotations: Cost-benefit Tradeoffs

Legacy GP elements already include input and output parameter descriptions that are defined in relation to a small set of possible types (e.g. string, integer, business object). Minimally these types can automatically be converted into corresponding ontology elements. At the same time our framework allows users to optionally refine these descriptions and to map service parameters onto more specific classes. For instance, rather than specifying an input parameter as a string, one might define it as an employee_name, which itself may be defined as a subclass of string in a domain specific ontology. While optional, more detailed descriptions enable more sophisticated reasoning functionality thereby leading to more and better support for the user.

There are however cost-benefit tradeoffs associated with the development of rich ontologies and annotations and it would be unrealistic to assume their existence from day one. Instead our expectation is that over time users will learn to appreciate the better support provided by these annotations and will be more willing and able to invest the necessary effort to develop them. Our mixed initiative framework does not assume the existence of rich and accurate ontologies and annotations. Clearly in the absence of such annotations, the support provided by our framework is not as powerful and may occasionally be itself inaccurate. It is therefore critical for this support to never hinder the user but rather to let the user choose when to invoke it and whether or not to follow its recommendations. As users invoke mixed initiative functionality and identify what appear to be inaccurate or incomplete annotations, it is critical to enable them to easily examine and, if necessary, modify these annotations (subject to proper approval procedures). As annotations become more complete and accurate, we expect GP users to increasingly rely on our mixed initiative support and to make fewer errors as they build composite applications and services (e.g. fewer mismatches between input and output parameters, fewer step omissions in the construction of composite processes, etc.). This in turn should translate into higher quality processes and an overall increase in productivity.

## 4. Overall Architecture & Underlying Reasoning

**Overall Architecture**

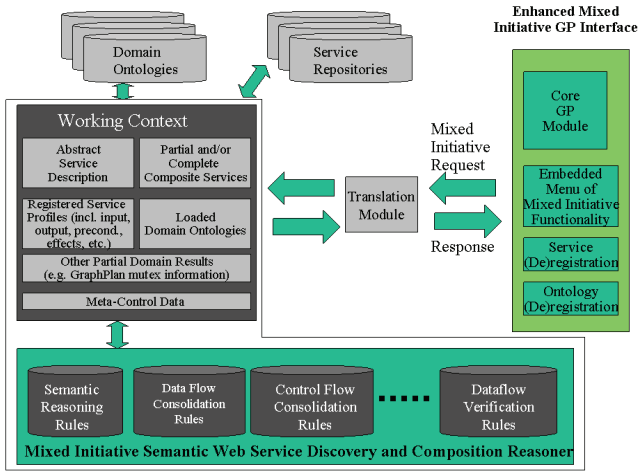

**Figure 4.** Overall architecture

The implementation of our mixed initiative semantic web service discovery and composition framework in the context of SAP's Guided Procedures comprises (figure 4):

1. Enhancements of the GP graphical user interface with access not just to core GP functionality (e.g. editing callable objects, actions and blocks) but also to a growing collection of mixed initiative service discovery and composition functionality. Invoking this mixed initiative functionality results in requests being sent to a mixed initiative semantic web service discovery and composition reasoner.
2. Services to (de)register both services and ontologies
3. The mixed initiative semantic web service and discovery reasoner itself, which is implemented as an independent module. As already indicated, while an initial version of this module has been integrated in GP, the module itself has been designed so that it could play a similar role in other service composition/process development environments

This latter module is implemented in the form of a rule-based engine (currently using JESS, a high-performance Java-based rule engine [7]). Rules in the engine implement a growing collection of mixed initiative service discovery and composition functionality, which itself combines two forms of reasoning:

1. **semantic reasoning** (e.g. reasoning about classes and subclasses as well as about more complex constructs supported by the OWL language)
2. **service composition planning** functionality implementing extensions of the highly efficient GraphPlan algorithm [10,11] – itself reimplemented using JESS rules.

This underlying reasoning functionality is further discussed in Subsections 4.2 and 4.3. Facts in the rule-based reasoner are organized in a working context (Figure 4). They include:

- An abstract description of the desired composite service
- A description of partial or complete service(s) generated to satisfy the user's request – these composite services may also include inconsistencies
- Profiles describing registered composable elements (or "services")
- Facts contained in or inferred from registered ontologies
- Partial domain results, produced while processing mixed initiative requests. This information, while dynamic, is maintained as it tends to change only slightly from one user request to the next (during the composition of a given service). Housekpeeing rules, not depicted in Figure 4, help remove facts that have been invalidated. Examples of partial results include nodes, edges, levels and "mutex" information derived as part of the Graphplan algorithm (see 4.3) or candidate matches for dataflow consolidation between two consecutive services.
- Meta-control data is also maintained in the working context in the form of predicates corresponding to different mixed initiative requests. These facts in turn trigger rules associated with the corresponding mixed initiative functionality, e.g rules implementing service discovery, parameter consolidation, dataflow verification, etc.

**Semantic Reasoning**

This functionality enables our module to load OWL ontologies and annotations and reason about them. This is done using an OWL-Lite Meta-Model, expressed in CLIPS, the modeling language used by JESS. An example of such a meta-model can be found in [22]. A translator is used to convert OWL-Lite ontologies into JESS triples. Our current implementation is based on Jena's RDF/XML Parser, ARP [25].

**Service composition planning**

This functionality is implemented using extensions of the GraphPlan algorithm. This is an algorithm that combines:

- reachability analysis to determine whether a given state (e.g. a combination of effects) can be reached

- from another state (e.g. the state reached after invoking an initial set of services), and
- disjunctive refinement, namely the addition of constraints between steps to resolve possible inconsistencies

In this algorithm, services and propositions (i.e. input, output, preconditions and effects in our model) are organized in layers in a "graphplan" that is iteratively analyzed and refined to obtain one or more service composition plans – if such plans exist. The graphplan consists of nodes, edges and layers (or levels). Possible inconsistencies are represented in the form of "mutex" information. This information in turn can be used to support mixed initiative functionality such as recommending possible ways in which to sequence services ("control flow"). Clearly, when used in one step, the GraphPlan algorithm can help identify all possible composite services satisfying an abstract description. Instead, we use a variation of this algorithm that enables us to find one or more plans at a time. This approach allows users to specify how many composite services they want to evaluate at a time and is also more practical, given the potentially large computational effort involved in identifying all possible composite services compatible with a given request. Other examples of mixed initiative functionality supported by this planning algorithm include:

- Identifying some or all services capable of producing a given effect or a given output
- Identifying all services that could be invoked following the execution of a given service
- Detecting conflicts between a selected service and other services already selected as part of a partial solution and suggesting ways of resolving these conflicts (using mutex information)
- Suggesting modifications to the abstract description of a desired composite service if no plan can be found for the current description

Graphplan expansion and the mutex generation are implemented as Jess rules, while plan extraction is implemented as a combination of Jess queries and Java functions.

## 5. Guided Procedure Scenario Revisited

In a typical interaction with the semantically enhanced version of GP, a user will provide a high level description of a desired composite service. This description can be entered using a wizard that allows users to specify desired service profile attributes (e.g. input/output parameters, preconditions and effects) in relation to loaded ontologies (e.g. see screen shot in Figure 5). This specification is loaded into the semantic service discovery and composition reasoner's working context, where it will help constrain future mixed initiative requests from the user. A simple (and admittedly naïve) request might be to automatically search for one or more composite services that match the user's composite service description. Other more typical requests are in the form of incremental steps, where users iteratively look for composable elements that help satisfy part of the service description, refine the control flow and data flow of selected composable elements, and possibly revise the original composite service until a satisfactory solution is obtained.

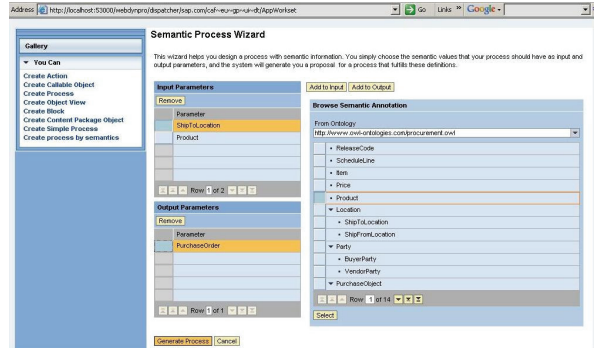

**Figure 5:** Specifying an abstract composite service profile in relation to concepts in an ontology

Figure 6 displays a typical screen shot, where a user invokes mixed initiative functionality to obtain suggestions on how to consolidate the input and output of two consecutive services intended to be part of a composite process referred to as "Purchase Order Scenario". Here, based on sub-class relationships in a domain ontology, the system recommends consolidating an output parameter called "warehouse address" with the "ship to location" input parameter of a subsequent service.

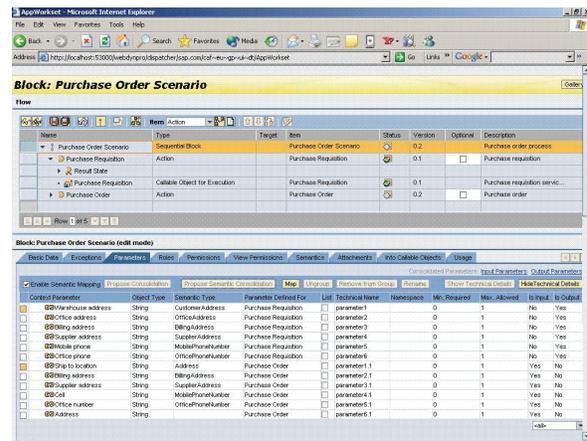

**Figure 6:** Suggestions on how to consolidate the input and output parameters of two consecutive services

## 6. Implementation Details and Evaluation

Our mixed initiative semantic service discovery and composition reasoner has been implemented using Jess.

Ontologies are expressed in OWL, while the services are described using a slightly modified fragment of OWL-S. An OWL metamodel [22] is loaded into Jess as facts. We use Jena to translate OWL documents into triples – also represented in Jess facts. Mixed initiative rules based on the GraphPlan algorithm have been implemented to support an initial set of mixed initiative functionality, including service discovery, dataflow consolidation, control flow and verification functionality.

The resulting system has been integrated with SAP's Guided Procedure framework and evaluated on an IBM laptop with a 1.80GHz Pentium M CPU and 1.50GB of RAM. The laptop was running Windows XP Professional OS, Java SDK 1.4.1 and Jess 6.1. Below, we report results obtained using ontologies from the Lehigh University Benchmark (LUBM) [26]. The results are based on the university example with around 50000 triples. Results are reported for repositories of 100, 500 and 1000 randomly generated semantic web services. Each randomly generated service had up to 5 inputs and 5 outputs. Input and output parameter types were randomly selected from the classes in the domain ontology. Performance, measured in terms of CPU times (in milliseconds), has been broken down as follows:

- Service and ontology loading time – this is typically done once when launching the system. Registering a single new service is an incremental process that only requires a tiny fraction of this time.
- Semantic reasoning time, which mainly involves completing domain ontologies once they have been loaded, is also typically performed just when launching the system
- Request processing: This is the time required to automatically generate composite services that match a randomly generated abstract composite service description. This time depends on the number of valid composite services one wishes to generate. For each service repository size, performance for two such values (number between parentheses) is reported.

As can be seen, the time it takes to produce multiple composite services ranges between 0.5 and 4 seconds. This seems quite acceptable, especially given that most of the time users will submit more incremental, and hence less time consuming, requests. The time it takes to load the system is higher than we would like, though we believe that, with some code optimization and possibly more powerful hardware, it will prove to be quite acceptable as well.

While encouraging, these are only preliminary results and further testing is needed to fully evaluate the scalability of our approach. In addition, detailed experimentation with actual users will be needed to fine tune the way in which mixed initiative functionality is presented and to eventually evaluate the full benefits of our approach from a productivity and solution quality standpoint.

| | *CPU time (in milliseconds)* | | |
|---|---|---|---|
| *Nb. Services (Nb. Sol.)* | Ontology and service loading | Semantic Reasoning | Request Processing |
| *100 (12)* | 54468 | 86475 | 1041 |
| *100 (211)* | 52445 | 89035 | 3141 |
| *500 (2)* | 52465 | 206687 | 511 |
| *500 (40)* | 53166 | 220227 | 1702 |
| *1000 (3)* | 54689 | 477467 | 1235 |
| *1000(78)* | 57944 | 457207 | 4116 |

## 7. Summary and Concluding Remarks

In this article, we have summarized ongoing work on the development of a mixed initiative semantic web service discovery and composition framework. In contrast to most work on semantic web service discovery and composition, our approach does not assume the existence of rich and accurate annotations from day one. Instead, it is intended to selectively intervene and assist users in some of their tasks by providing suggestions, identifying inconsistencies, and completing some of the user's decisions. Users are always in control and decide when and how much to delegate to supporting functionality. The quality and accuracy of the support provided by our framework is intended to improve over time, as users learn to develop richer and more accurate annotations.

An initial version of this framework has been integrated and evaluated in the context of SAP's Guided Procedures, a central element of the company's Enterprise Service Architecture. Initial empirical results have confirmed the viability of our underlying reasoning framework, which leverages a combination of semantic reasoning functionality and service composition planning functionality based on the GraphPlan algorithm Rather than being implemented in a monolithic manner, this functionality has been broken down and extended to support an initial collection of mixed initiative functionality. Over time, we plan to further extend and refine the way in which this functionality is presented to users. While our initial results are promising, we recognize that additional testing (and fine tuning) will be needed to fully realize and evaluate the potential of our approach and to measure actual improvements in user productivity and solution quality.

## Acknowledgements

The "Smart Service Discovery and Composition" work reported herein has been conducted as a collaboration between Sadeh Consulting and SAP Inspire The authors would like to thank Lutz Heuser and Claudia Alsdorf for supporting this project. Special thanks to Shuyuan Chen, Horst Werner, Kiril Bratanov and Daniel Hutzel for their